\documentclass[12pt,a4paper,fleqn]{article}

\usepackage[DIV13]{typearea}
\usepackage{amsmath}
\usepackage{amssymb}
\usepackage{amsfonts}
\usepackage{amscd}
\usepackage{graphicx}
\usepackage[footnotesize]{caption2}
\usepackage{cite}

\newcommand{\CiteSeeSaw}{\cite{Yanagida:1980,Glashow:1979vf,Gell-Mann:1980vs,Mohapatra:1980ia}}

\DeclareMathOperator{\Tr}{Tr}
\DeclareMathOperator{\diag}{diag}

\newcommand{\GeV}{\ensuremath{\:\mathrm{GeV}}}

\newcommand{\Eqref}[1]{eq.~\eqref{#1}}
\newcommand{\Ineqref}[1]{ineq.~\eqref{#1}}
\newcommand{\Figref}[1]{figure \ref{#1}}
\newcommand{\Tabref}[1]{table \ref{#1}}
\newcommand{\Secref}[1]{section \ref{#1}}

\def\D{\mathrm{d}}
\def\I{i}

\newcommand{\Dmsol}{\Delta m^2_{\smash[b]{\scriptscriptstyle\odot}}}
\newcommand{\Dmatm}{\Delta m^2_\mathrm{a}}

\allowdisplaybreaks[1]

\begin{document}

\begin{titlepage}
\renewcommand{\thefootnote}{\alph{footnote}}

\vspace*{-2.0cm}
\begin{flushright}
TUM-HEP-550/04\\
\end{flushright}

\vspace*{1.0cm}

\renewcommand{\thefootnote}{\fnsymbol{footnote}}

{\begin{center}
{\Large\bf Stability of Texture Zeros under Radiative Corrections in
See-Saw Models}
\end{center}}
\renewcommand{\thefootnote}{\alph{footnote}}

\vspace*{.8cm}

\begin{center} 
{
\textbf{
  Claudia Hagedorn\footnote[1]{\makebox[1.cm]{Email:} chagedor@ph.tum.de},
  J\"{o}rn Kersten\footnote[2]{\makebox[1.cm]{Email:} jkersten@ph.tum.de},
  and
  Manfred Lindner\footnote[3]{\makebox[1.cm]{Email:} lindner@ph.tum.de}}
}
\end{center}

\begin{center}
Theoretische Physik, Physik-Department,
Technische Universit\"at M\"unchen,\\
James-Franck-Stra{\ss}e, 85748~Garching bei M{\"u}nchen, Germany
\end{center}

\vspace*{0cm}

\begin{abstract}
\noindent
It has been shown that only certain neutrino mass matrices with 
texture zeros are compatible with existing data. We discuss the 
stability of phenomenological consequences of texture zeros  
under radiative corrections in the type-I see-saw scenario. We 
show that under certain conditions additional patterns are allowed 
due to these effects. 
\end{abstract}

\end{titlepage}

\newpage

\section{Introduction}
Remarkable progress has been made in recent years in the determination
of neutrino masses and mixings, especially due to the results from
oscillation experiments. The fact that two of the three mixings 
turned out to be large came as a surprise and shows that we 
are still far from a fundamental understanding of the flavour 
structure of the Standard Model (SM) and its extensions which 
include neutrino masses. This surprise should remind us that the 
absolute neutrino masses need not be hierarchical as na\"{i}vely 
expected from the analogy to the quark sector. This is important,
since renormalization group (RG) effects can then have a considerable
impact on the evolution of flavour structures towards the scale where 
they are explained.

Similar to models in the quark sector 
\cite{Fritzsch:1978vd,Georgi:1979df,Harvey:1980je,Dimopoulos:1992za,Giudice:1992an,He:1990eh,Ramond:1993kv,Zhou:2003ji,Xing:2003yj},
one framework to understand the neutrino masses and mixings is to 
assume certain elements of the mass matrix to be zero
\cite{Frampton:2002yf,Guo:2002ei,Kageyama:2002zw,Bando:2003ei,%
Honda:2003pg,Xing:2003ic,Xing:2003ez,Bando:2003wb,Desai:2002sz,%
Xing:2003zd,Zhou:2004wz,Randhawa:2002jt,Koide:2002cj,Gill:1998qa}. 
Such texture zeros can be produced, for example, by flavour symmetries,
see e.g.\ \cite{Grimus:2004hf}.  However, these symmetries
are often broken at a high energy scale, so that 
RG running destroys the texture zeros.
A classification of textures with two or more zeros and their 
compatibility with experimental data was given in refs.~\cite{Frampton:2002yf,Guo:2002ei}.
We will study how RG corrections affect these results.  
In particular, we will show that some of the patterns which have 
been excluded become allowed, 
if the running leads to sufficiently large entries in the positions 
of the former zeros.

The running of neutrino parameters is described at low energies
by the RG equations for the leading neutrino mass operator.
In the SM and its minimal supersymmetric extension (MSSM) with massive
neutrinos, this is a dimension 5 operator, 
whose energy dependence is determined by the RG equation
\cite{Chankowski:1993tx,Babu:1993qv,Antusch:2001ck,Antusch:2001vn}
\begin{equation} \label{eq:BetaKappa}
 16\pi^2 \, \frac{\D\kappa}{\D t} =
 C\,(Y_e^\dagger Y_e)^T\,\kappa+C\,\kappa\,(Y_e^\dagger Y_e) + \alpha\,\kappa\;.
\end{equation}
The neutrino mass matrix is defined as $m_\nu=\frac{v^2}{4} \kappa$, 
where $v \approx 246$~GeV in the SM and $v \approx \sin\beta \cdot 246$~GeV 
in the MSSM. The renormalization scale $\mu$ enters as usual via 
$t=\ln(\mu/\mu_0)$, and
$C = -\frac{3}{2}$ in the SM, $C = 1$ in the MSSM.
Finally, $\alpha$ is a flavour scalar containing e.g.\ gauge couplings. Hence, 
the last term in the RG equation is flavour-blind and leads only to an 
overall rescaling which does not affect the mixing parameters.

In this paper, we work in the basis where the Yukawa matrix of the charged 
leptons $Y_e$ is diagonal, i.e.\ we use the same classification of textures 
as in \cite{Frampton:2002yf,Guo:2002ei}. From 
\Eqref{eq:BetaKappa} it follows then immediately that the radiative corrections 
to each element of the neutrino mass matrix (or, equivalently, $\kappa$) 
are proportional to this element itself, so that a zero entry remains zero.  
However, this changes in the type-I see-saw scenario \CiteSeeSaw{} at 
energies larger than the masses of the heavy singlet neutrinos that are 
introduced in order to explain the smallness of the light neutrino masses.  
Above this threshold, the neutrino Yukawa
couplings $Y_\nu$ contribute to the RG equations.  In general, they
are non-diagonal and hence cause a mixing of the mass matrix elements,
so that the elements which are zero at the high-energy scale obtain a
finite value at low energies and thus texture zeros are destroyed.  
We will determine under which conditions this modifies the compatibility
of textures with experimental data.  We will apply
a bottom-up approach, i.e.\ we will study if the RG running from low to
high energies can lead to a zero element.  Therefore, we will refer to
this possibility as radiative generation rather than destruction of
texture zeros.

\section{Conditions for the Stability of Textures} 
\label{sec:ConditionsForStability}
\subsection{Running of the Effective Neutrino Mass Matrix}
In the full theory with singlet neutrinos, i.e.\ above
the highest see-saw scale, the relevant quantity for our discussion
is the effective light neutrino mass matrix defined as
\begin{equation}
        m_\nu = -\frac{v^2}{2} Y_\nu^T M^{-1} Y_\nu \;,
\end{equation}
where $M$ is the mass matrix of the heavy singlets and $v$ is the
relevant Higgs vacuum expectation value at low energy.  From the RG equations for $Y_\nu$ and $M$
\cite{Machacek:1984fi,Pirogov:1998tj,Falck:1986aa,Tanimoto:1995bf,Casas:1999tp,Casas:1999ac}
one obtains
\begin{equation} \label{eq:Betamnu}
        16\pi^2 \, \frac{\D m_\nu}{\D t} =
        (C_e \, Y_e^\dagger Y_e + C_\nu \, Y_\nu^\dagger Y_\nu)^T \, m_\nu +
        m_\nu \, (C_e \, Y_e^\dagger Y_e + C_\nu \, Y_\nu^\dagger Y_\nu) + 
        \alpha' \, m_\nu
\end{equation}
with
$C_e = -\frac{3}{2}$, $C_\nu=\frac{1}{2}$ in the SM,
$C_e = C_\nu = 1$ in the MSSM,
and
\begin{subequations}
\begin{align}
        \alpha'_\mathrm{SM} =&
        -\frac{9}{10} g_1^2 -\frac{9}{2} g_2^2 + 2\Tr(Y_\nu^\dagger Y_\nu) +
        2 \, (y_\tau^2 + y_\mu^2 + y_e^2) + {}
\nonumber\\
& {} +
        6 \, (y_t^2 + y_b^2 + y_c^2 + y_s^2 + y_d^2 + y_u^2) \;,
\\
        \alpha'_\mathrm{MSSM} =&
        -\frac{6}{5} g_1^2 - 6 g_2^2 + 2 \Tr(Y_\nu^\dagger Y_\nu) +
        6 \, (y_t^2 + y_c^2 + y_u^2) \;,
\end{align}
\end{subequations}
where $y_i$ are Yukawa couplings.
We use GUT charge normalization for the U(1) gauge coupling.
Between the mass thresholds, the running is modified
because the singlets are successively integrated out \cite{King:2000hk}.
Note that the RG equations for the effective mass operators
cannot be obtained from \Eqref{eq:Betamnu} by omitting suitable
Yukawa couplings in general \cite{Antusch:2002rr}.

\subsection{Mass Matrix Elements} \label{sec:MassMatrixElements}
In the bottom-up approach, one can understand under which circumstances
the radiative generation of texture zeros is possible.  If an element
of the neutrino mass matrix, say $m_{\nu_{ij}}$, is to be zero at the
GUT scale $M_\mathrm{GUT}$, the sum of its value at the see-saw scale
$M_1$ and its change due to the running between this scale and
$M_\mathrm{GUT}$ has to be zero.  In linear approximation, this means
\begin{equation} \label{eq:Start}
        \left. m_{\nu_{ij}} \right|_{M_1} \approx
        \left. -\dot m_{\nu_{ij}} \right|_{M_1} \cdot
         \ln\frac{M_\mathrm{GUT}}{M_1} \;.
\end{equation}
If $m_{\nu_{ij}}$ is complex, this relation must be satisfied for both 
its real and its imaginary part. Hence, we can assume $m_\nu$ to be real
in the following derivation if we keep in mind that each equation
actually stands for two.  We also assume that all singlet neutrinos are
integrated out at the mass $M_1$ of the lightest singlet.  To obtain a
conservative bound, we choose this mass to be larger than about
$10^9$ GeV.  Of course, smaller values are possible in principle, but
then the neutrino Yukawa couplings have to be rather small as well, so
that the RG evolution is actually suppressed.  In the numerical examples
presented later on, the see-saw scale lies comfortably above this lower
bound.  With $M_1 \gtrsim 10^9$ GeV and $M_\mathrm{GUT} \sim 10^{16}$ GeV, \Eqref{eq:Start} becomes
\begin{equation} \label{eq:StartLimit}
        -\frac{16\pi^2 \, \dot m_{\nu_{ij}}}{m_{\nu_{ij}}}
        \gtrsim 10 \;.
\end{equation}
Here and in the following, we implicitly assume the values of
energy-dependent quantities like $m_\nu$ to be taken at $M_1$.
Keeping only the top and tau Yukawa couplings and introducing the
abbreviation $H := Y_\nu^\dagger Y_\nu$, the RG equation
\eqref{eq:Betamnu} can be rewritten for an individual matrix
element,
\begin{equation} \label{eq:BetamnuijMSSM}
        16\pi^2 \, \dot m_{\nu_{ij}} =
        (m_\nu H)_{ij} + (m_\nu H)_{ji} + \sigma \, m_{\nu_{ij}}
\end{equation}
with
\begin{equation}
        \sigma := 2 \Tr H + (\delta_{i3}+\delta_{j3}) \, y_\tau^2 +
         6 y_t^2 - \frac{6}{5} g_1^2 - 6 g_2^2 
        \stackrel{M_1=10^{13}\GeV}{\approx}
        2 \Tr H + (\delta_{i3}+\delta_{j3}) \, y_\tau^2 + 1
\end{equation}
in the MSSM.  Note that $\sigma$ is
always positive.  Plugging \eqref{eq:BetamnuijMSSM} into
\Ineqref{eq:StartLimit} and collecting all terms independent of $m_\nu$
on the r.h.s.\ yields
\begin{equation} \label{eq:Master}
        -\frac{\sum_{k \neq j} m_{\nu_{ik}} H_{kj} +
                   \sum_{k \neq i} m_{\nu_{jk}} H_{ki}}{m_{\nu_{ij}}}
        \gtrsim 10 + \sigma + H_{ii} + H_{jj} \;.
\end{equation}
Careful inspection of this relation shows that
radiative generation of texture zeros requires at least one element of
$m_\nu$ to be roughly an order of magnitude larger than the value of the
zero candidate at the see-saw scale.  In the complex case, both the real
and the imaginary parts of the matrix elements have to satisfy this
requirement.

To see this, consider the easiest case first: let us assume that the
l.h.s.\ of \Ineqref{eq:Master} is dominated by a single
term, say $m_{\nu_{ik}} H_{kj}$ ($i \neq j$), and that $H_{kj} \sim 1$.
Then $H_{jj} \sim 1$, $H_{ii}$ is small, and $\Tr H \sim 2$.  Hence, we
obtain
\begin{equation}
        -\frac{m_{\nu_{ik}}}{m_{\nu_{ij}}} \gtrsim 16 \;.
\end{equation}
In principle, $H_{kj}$ could be as large as 3 as long as all 
$|Y_{\nu_{ij}}| \lesssim 1$.  This would reduce the required hierarchy
somewhat, but the difference is not dramatic as the terms containing $H$
on the r.h.s.\ become larger as well.

The smallest hierarchy is expected if all elements of $H$ are large.
For $H_{ij} \sim 1$, we find
\begin{equation}
        -\frac{\sum_{k \neq j} m_{\nu_{ik}} +
                   \sum_{k \neq i} m_{\nu_{jk}}}{m_{\nu_{ij}}}
        \gtrsim 19 \;,
\end{equation}
so that
\begin{equation}
        -\frac{m_{\nu_{ik}}}{m_{\nu_{ij}}} \gtrsim 5
\end{equation}
if all elements of the neutrino mass matrix except $m_{\nu_{ij}}$ are
approximately equal.  Thus, the required hierarchy can be a bit smaller
than an order of magnitude.  However, in this case it is difficult to
find a Yukawa matrix which is perturbative and still reproduces
the correct neutrino mass parameters via the see-saw formula.
Hence, we conclude that the above statement about the required hierarchy
in $m_\nu$ is reasonably conservative.

In the SM, the first two terms on the r.h.s.\ of \Eqref{eq:BetamnuijMSSM} are multiplied
by an additional factor of $\frac{1}{2}$, and $\sigma$ is replaced by
\begin{equation}
        \sigma =
        2 \Tr H + 6 y_t^2 - \frac{9}{10} g_1^2 - \frac{9}{2} g_2^2
        \stackrel{M_1=10^{13}\GeV}{\approx}
        2 \Tr H + 0.4 \;,
\end{equation}
where we have neglected the tau Yukawa coupling, too, since it is
always small in the SM.  Because of the change in
\Eqref{eq:BetamnuijMSSM}, the term $10+\sigma$ in
\Ineqref{eq:Master} has to be multiplied by 2.  Thus, a hierarchy is
necessary in the SM as well, and it has to be even larger than in the
MSSM, so that the generation of texture zeros becomes harder.  On the
other hand, the effects of the running between the mass scales of the
heavy singlets can be especially significant in the SM
\cite{Antusch:2002rr} and might provide a loophole attenuating the
restrictions found in this section.

\subsection{Mass Eigenvalues}
Let us now see what the hierarchy requirement for the elements of the
neutrino mass matrix means for the mass eigenvalues.  For this purpose,
we express the matrix elements in terms of the eigenvalues
$m_1,m_2,m_3$ and the mixing parameters.  We
use the approximations $\theta_{13} \approx 0$
and $\Dmsol\ll\Dmatm$.  For a normal mass hierarchy, we have $m_1 := m$,
$m_2 = \sqrt{m^2+\Dmsol}$ and $m_3 \approx \sqrt{m^2+\Dmatm}$.  Using
\begin{equation}
        U^T m_\nu \, U = \diag(m_1,m_2,m_3)
\end{equation}
and the standard parameterization for the lepton mixing matrix $U$ in the 
limit $\theta_{13}=0$, 
\begin{equation} \label{eq:StandardParametrizationU}
 U = \begin{pmatrix}
 c_{12} & s_{12} & 0 \\
 -c_{23}s_{12} & c_{23}c_{12} & s_{23} \\
 s_{23}s_{12} & -s_{23}c_{12} & c_{23}
 \end{pmatrix}
 \cdot \diag(e^{-\I\varphi_1/2},e^{-\I\varphi_2/2},1) \;,
\end{equation}
we obtain (assuming without loss of generality that all unphysical
phases in $U$ are zero)
\begin{subequations} \label{eq:mnuij}
\begin{align}
        m_{\nu_{11}} &\approx
         \sqrt{m^2+\Dmsol} \,s_{12}^2 \,e^{\I\varphi_2} +
         m \,c_{12}^2 \,e^{\I\varphi_1} \;, \\
        m_{\nu_{22}} &\approx
         \sqrt{m^2+\Dmatm} \,s_{23}^2 +
         \sqrt{m^2+\Dmsol} \,c_{12}^2 c_{23}^2 \,e^{\I\varphi_2} +
         m \,s_{12}^2 c_{23}^2 \,e^{\I\varphi_1} \;, \\
        m_{\nu_{33}} &\approx
         \sqrt{m^2+\Dmatm} \,c_{23}^2 +
         \sqrt{m^2+\Dmsol} \,c_{12}^2 s_{23}^2 \,e^{\I\varphi_2} +
         m \,s_{12}^2 s_{23}^2 \,e^{\I\varphi_1} \;, \\
        m_{\nu_{12}} &\approx
         \Bigl( \sqrt{m^2+\Dmsol} \,e^{\I\varphi_2} - m \,e^{\I\varphi_1}
         \Bigr) \, c_{12} s_{12} c_{23} \;, \\
        m_{\nu_{13}} &\approx
         -\Bigl( \sqrt{m^2+\Dmsol} \,e^{\I\varphi_2} - m \,e^{\I\varphi_1}
         \Bigr) \, c_{12} s_{12} s_{23} \;, \\
        m_{\nu_{23}} &\approx
         \Bigl( \sqrt{m^2+\Dmatm} - \sqrt{m^2+\Dmsol} \,c_{12}^2 \,e^{\I\varphi_2} -
         m \,s_{12}^2 \,e^{\I\varphi_1} \Bigr) \, c_{23} s_{23} \;,
\end{align}
\end{subequations}
where $s_{12} := \sin\theta_{12}$, $c_{12} := \cos\theta_{12}$ etc.
If the mass spectrum is hierarchical, i.e.\ $m \approx 0$, the elements
of the first row and column of $m_\nu$ are of the order of
$\sqrt{\Dmsol}$, while the others are of the order of $\sqrt{\Dmatm}$.
Hence, the maximal hierarchy between two elements of $m_\nu$ is roughly
$\sqrt{\Dmsol/\Dmatm} \sim 1/6$, which is not enough for the generation
of a texture zero.  This statement holds independent of the Majorana
phases $\varphi_1$ and $\varphi_2$, since each matrix element is
dominated by just a single term.  Note that the running of the mixing
angles between the electroweak and the see-saw scale does not change
this conclusion, since it is not significant for a strong normal
hierarchy \cite{Chankowski:1999xc,Casas:1999tg,Antusch:2003kp}.

For quasi-degenerate neutrinos, i.e.\ $m^2\gg\Dmatm$, the above formulae
simplify to
\begin{subequations} \label{eq:mnuijDegen}
\begin{align}
        m_{\nu_{11}} &\approx
         m \left( s_{12}^2 \,e^{\I\varphi_2} + c_{12}^2 \,e^{\I\varphi_1}
         \right) , 
        \label{eq:mnu11Degen} \\
        m_{\nu_{22}} &\approx
         m \, \left( s_{23}^2 + c_{12}^2 c_{23}^2 \,e^{\I\varphi_2} +
         s_{12}^2 c_{23}^2 \,e^{\I\varphi_1} \right) , \\
        m_{\nu_{33}} &\approx
         m \left( c_{23}^2 + c_{12}^2 s_{23}^2 \,e^{\I\varphi_2} +
         s_{12}^2 s_{23}^2 \,e^{\I\varphi_1} \right) , \\
        m_{\nu_{12}} &\approx
         m \left( e^{\I\varphi_2} - e^{\I\varphi_1} \right)
         c_{12} s_{12} c_{23} \;, \\
        m_{\nu_{13}} &\approx
         -m \left( e^{\I\varphi_2} - e^{\I\varphi_1} \right)
         c_{12} s_{12} s_{23} \;, \\
        m_{\nu_{23}} &\approx
         m \left( 1 - c_{12}^2 \,e^{\I\varphi_2} -
         s_{12}^2 \,e^{\I\varphi_1} \right) c_{23} s_{23} \;.
\end{align}
\end{subequations}
As the situation now depends on the values of the Majorana phases, we
consider the real cases first.  For $\varphi_1=\varphi_2=0$, the diagonal
elements of the
mass matrix approach $m$, while the off-diagonal entries go to zero.
One can show  that $m_{\nu_{ij}}/m_{\nu_{ii}}$ ($i \neq j$) are
monotonous functions of $m$.  Hence, for a sufficiently large neutrino
mass scale the required hierarchy in the matrix elements arises and the
radiative generation of zeros in the off-diagonal elements becomes
possible.  In contrast, it is not possible to generate zeros in the
diagonal entries.

If one Majorana phase is $\pi$ and the other one zero, i.e.\ one of the
mass eigenstates has a negative CP parity, all
$|m_{\nu_{ij}}\!|$ become large for $m\to\infty$.  The smallest entry is
$|m_{\nu_{11}}\!| \approx m \cos2\theta_{12}$, but it cannot be much
smaller than the other matrix elements due to the experimental bound
$\cos2\theta_{12}>0.22$ at the $3\sigma$ level \cite{Maltoni:2003da}.
Quantum corrections between the electroweak and the see-saw scale are
not likely to change this situation, since they are not significant in
the SM, and since they only cause a decrease of $\theta_{12}$ in the
MSSM \cite{Miura:2002nz} unless $\theta_{13}$ is close to its
experimental upper limit \cite{Antusch:2003kp}.  Consequently, the
generation of a texture zero is unlikely.

The last real case, $\varphi_1=\varphi_2=\pi$, leads to the same
conclusions for the first row and column as $\varphi_1=\varphi_2=0$,
because only the phase difference is relevant here.  The situation
changes for the diagonal elements $m_{\nu_{22}}$ and $m_{\nu_{33}}$,
which go to $\pm m \cos2\theta_{23}$ now, so that zeros can be generated
in these elements if $\theta_{23}$ is close to $45^\circ$.  The
remaining off-diagonal entry
$m_{\nu_{23}}$ stays large for large atmospheric mixing and thus cannot
be driven to zero by the RG in this case.

For a complex mass matrix, i.e.\ arbitrary phases, the hierarchy
requirement must be satisfied by both the real and the imaginary parts
of the respective matrix elements, as mentioned in the beginning of this
section.  
In the 23-block of $m_\nu$, this is not possible even if the Majorana
phases are very small, because
the hierarchy in the real cases is due to a cancellation
between the three terms contributing to $m_{\nu_{23}}$,
only two of which contribute to the imaginary part.
Consequently, one expects at first sight that texture zeros 
can only be generated radiatively in the CP-conserving cases.
However, this is too pessimistic, since adding a small imaginary part to
the neutrino mass matrix does not change the mixing angles and mass
squared differences significantly, except for special cases such as
exactly degenerate mass eigenvalues.  This allows us to make the imaginary part
vanish by slightly adjusting the initial mass matrix, if the values of
the Majorana phases are not too far away from 0 or $\pi$.
For the elements of the first row and column, only the difference of the
phases is relevant.  Hence, the radiative generation of a texture zero
in these positions is possible for arbitrary Majorana phases, provided
that their difference is small.

If $\theta_{13}$ is non-zero, a finite Dirac phase $\delta$ could be an
obstacle for the generation of texture zeros as well.  However, as
$\theta_{13}$ is experimentally restricted to be relatively small and
does not grow too much under the RG \cite{Antusch:2003kp}, the
same argument holds as for small Majorana phases, so that the value of
$\delta$ is not very important.

In the case of an inverted mass hierarchy, we use the convention 
$m_3 := m$, $m_2 = \sqrt{m^2+|\Dmatm|}$ and 
$m_1 = \sqrt{m^2+|\Dmatm|-\Dmsol} \approx m_2$.
Then the elements of the neutrino mass matrix can be expressed as
\begin{subequations} \label{eq:mnuijInv}
\begin{align}
        m_{\nu_{11}} &\approx
         \sqrt{m^2+|\Dmatm|} \left( s_{12}^2 \,e^{\I\varphi_2} +
         c_{12}^2 \,e^{\I\varphi_1} \right) ,
        \label{eq:mnu11Inv} \\
        m_{\nu_{22}} &\approx
         \sqrt{m^2+|\Dmatm|} \,c_{23}^2 \left( c_{12}^2 \,e^{\I\varphi_2} +
         s_{12}^2 \,e^{\I\varphi_1} \right) + m \,s_{23}^2 \;, \\
        m_{\nu_{33}} &\approx
         \sqrt{m^2+|\Dmatm|} \,s_{23}^2 \left( c_{12}^2 \,e^{\I\varphi_2} +
         s_{12}^2 \,e^{\I\varphi_1} \right) + m \,c_{23}^2 \;, \\
        m_{\nu_{12}} &\approx
         \Bigl( \sqrt{m^2+|\Dmatm|} \,e^{\I\varphi_2} -
         \sqrt{m^2+|\Dmatm|-\Dmsol} \,e^{\I\varphi_1} \Bigr) \,
         c_{12} s_{12} c_{23} \;, \\
        m_{\nu_{13}} &\approx
         -\Bigl( \sqrt{m^2+|\Dmatm|} \,e^{\I\varphi_2} -
         \sqrt{m^2+|\Dmatm|-\Dmsol} \,e^{\I\varphi_1} \Bigr) \,
         c_{12} s_{12} s_{23} \;, \\
        m_{\nu_{23}} &\approx
         \Bigl( -\sqrt{m^2+|\Dmatm|} \left( c_{12}^2 \,e^{\I\varphi_2} +
         s_{12}^2 \,e^{\I\varphi_1} \right) + m \Bigr)
         \, c_{23} s_{23}
\end{align}
\end{subequations}
for $\theta_{13} \approx 0$.  We have not used the approximation 
$m_1 \approx m_2$ for $m_{\nu_{12}}$ and $m_{\nu_{13}}$ to avoid
underestimating the size of these entries for small $m$ and equal
Majorana phases.  For the other matrix elements, this approximation is
sufficiently accurate, since the mass eigenvalues are multiplied by
$s_{12}^2$ and $c_{12}^2$, respectively.  Experimentally, these numbers
are known to be unequal (and this does not change during the running up
to the see-saw scale, as mentioned above), so that there is no complete
cancellation if $\Dmsol$ is neglected.

A strong mass hierarchy causes most elements of $m_\nu$ to be
of the order of $\sqrt{\Dmatm}$, so that the radiative generation of
texture zeros in these positions is not possible.  The exceptions are
$m_{\nu_{12}}$ and $m_{\nu_{13}}$, which are smaller than the other
elements by a factor of about $\Dmsol/\Dmatm$ for equal Majorana phases.
Hence, these entries can become zero at the GUT scale even for
relatively small values of $m$.

For quasi-degenerate masses, the conclusions are the same as for a
normal mass ordering.  This is not surprising as the only difference is
the sign of the atmospheric mass squared difference, which is not
important for $|\Dmatm| \ll m^2$.

An overview of the results of this section is given in
\Tabref{tab:RadGenSummary}.  The positions in the neutrino mass matrix
where a texture zero can be generated radiatively are marked by a
``{\Large$\scriptstyle\circ$}''.
\begin{table}
\centering
\renewcommand{\arraystretch}{1.8}
\begin{tabular}{llc}
\hline \\[-5.5ex]
Neutrino masses & Majorana phases & \\
\hline
Normal hierarchy, $m_1 \approx 0$ & arbitrary &
 \Large $\left(\begin{smallmatrix}
 \makebox[.8ex]{\rule[.1ex]{0ex}{.6ex}$\scriptstyle\cdot$} &\cdot &\cdot
 \\
 \cdot &\makebox[.8ex]{\rule[.1ex]{0ex}{.6ex}$\scriptstyle\cdot$} &\cdot
 \\
 \cdot &\cdot &\makebox[.8ex]{\rule[.1ex]{0ex}{.6ex}$\scriptstyle\cdot$}
 \end{smallmatrix}\right)$
\\
\hline
Inverted hierarchy, $m_3 \approx 0$ & $\varphi_1\approx\varphi_2$ &
 \Large $\left(\begin{smallmatrix}
 \cdot & \circ & \circ \\
 \circ & \cdot & \cdot \\
 \circ & \cdot & \cdot
 \end{smallmatrix}\right)$
\\
& $\varphi_1\not\approx\varphi_2$ &
 \Large $\left(\begin{smallmatrix}
 \makebox[.8ex]{\rule[.1ex]{0ex}{.6ex}$\scriptstyle\cdot$} &\cdot &\cdot
 \\
 \cdot &\makebox[.8ex]{\rule[.1ex]{0ex}{.6ex}$\scriptstyle\cdot$} &\cdot
 \\
 \cdot &\cdot &\makebox[.8ex]{\rule[.1ex]{0ex}{.6ex}$\scriptstyle\cdot$}
 \end{smallmatrix}\right)$
\\
\hline
Quasi-degenerate & $\varphi_1\approx\varphi_2\approx0$ &
 \Large $\left(\begin{smallmatrix}
 \cdot & \circ & \circ \\
 \circ & \cdot & \circ \\
 \circ & \circ & \cdot
 \end{smallmatrix}\right)$
\\
& $\varphi_{1,2}\approx0$, $\varphi_{2,1}\approx\pi$ &
 \Large $\left(\begin{smallmatrix}
 \makebox[.8ex]{\rule[.1ex]{0ex}{.6ex}$\scriptstyle\cdot$} &\cdot &\cdot
 \\
 \cdot &\makebox[.8ex]{\rule[.1ex]{0ex}{.6ex}$\scriptstyle\cdot$} &\cdot
 \\
 \cdot &\cdot &\makebox[.8ex]{\rule[.1ex]{0ex}{.6ex}$\scriptstyle\cdot$}
 \end{smallmatrix}\right)$
\\
& $\varphi_1\approx\varphi_2\approx\pi$ &
 \Large $\left(\begin{smallmatrix}
 \cdot & \circ & \circ \\
 \circ & \circ & \cdot \\
 \circ & \cdot & \circ
 \end{smallmatrix}\right)$
\\
& $\varphi_1\approx\varphi_2\not\approx0,\pi$ &
 \Large $\left(\begin{smallmatrix}
 \cdot & \circ & \circ \\
 \circ & \cdot & \cdot \\
 \circ & \cdot & \cdot
 \end{smallmatrix}\right)$
\\
& $\varphi_1\not\approx\varphi_2$ &
 \Large $\left(\begin{smallmatrix}
 \makebox[.8ex]{\rule[.1ex]{0ex}{.6ex}$\scriptstyle\cdot$} &\cdot &\cdot
 \\
 \cdot &\makebox[.8ex]{\rule[.1ex]{0ex}{.6ex}$\scriptstyle\cdot$} &\cdot
 \\
 \cdot &\cdot &\makebox[.8ex]{\rule[.1ex]{0ex}{.6ex}$\scriptstyle\cdot$}
 \end{smallmatrix}\right)$
\\
\hline
\end{tabular}
\caption{Possible positions of radiatively generated texture zeros in
 the neutrino mass matrix, marked by a ``{\Large$\scriptstyle\circ$}''.
 For $\varphi_1\approx\varphi_2\approx\pi$, at most 3 of the 4 zeros can
 be produced at the same time.
}
\label{tab:RadGenSummary}
\end{table}
Comparing these results with the classification of two-zero textures in
the literature, we find that three of the six forbidden textures in
\cite{Guo:2002ei} can be reconciled with data by the RG evolution.
These are the patterns of class F,
\[
        \begin{pmatrix}
        \times & 0 & 0 \\ 0 & \times & \times \\ 0 & \times & \times
        \end{pmatrix} ,
        \begin{pmatrix}
        \times & 0 & \times \\ 0 & \times & 0 \\ \times & 0 & \times
        \end{pmatrix} ,
        \begin{pmatrix}
        \times & \times & 0 \\ \times & \times & 0 \\ 0 & 0 & \times
        \end{pmatrix} ,
\]
where the crosses ``$\times$'' stand for the non-zero entries.
On the other hand, class E, i.e.\ the matrices of the form
\[
        \begin{pmatrix}
        0 & \times & \times \\ \times & 0 & \times \\ \times &\times &\times
        \end{pmatrix} ,
        \begin{pmatrix}
        0 & \times & \times \\ \times & \times & \times \\ \times &\times &0
        \end{pmatrix} ,
        \begin{pmatrix}
        0 & \times & \times \\ \times & \times & 0 \\ \times & 0 & \times
        \end{pmatrix} ,
\]
remain forbidden.  The reason for this is the zero in the 11-element,
which cannot be generated radiatively due to the large deviation of the
solar mixing angle from $\pi/4$ (cf.\ eqs.\ \eqref{eq:mnu11Degen},
\eqref{eq:mnu11Inv}).  One could hope to circumvent this
problem by assuming that this texture zero exists already at low
energies and that only the second one is created by the running.
However, this is not possible, since $m_{\nu_{11}}=0$ requires either a
strong hierarchy or a difference of $\pi$ between the Majorana phases,
both of which prevent the generation of another texture zero.

Thus, the number of two-zero textures which are at least marginally
compatible with experimental data can be raised from 9 to 12 if one
includes RG effects.  Furthermore, \Tabref{tab:RadGenSummary} shows that a
number of textures with three zeros, none of which is allowed at low
energies \cite{Frampton:2002yf}, should be possible as well.

\section{Examples for Radiatively Created Texture Zeros}
In this section, we give numerical examples for the radiative generation
of texture zeros.  The RG evolution from the GUT scale to low energies
is calculated numerically, starting with the desired texture zeros in
the neutrino mass matrix and suitable values for the other parameters
which ensure that all low-energy observables are compatible with
experiment.  This time, we solve the complete set of coupled
differential equations and take into account the threshold corrections
arising when the singlet neutrinos are successively integrated out at
different energies \cite{King:2000hk,Antusch:2002rr}.  
The examples are intended as a proof of
principle, and therefore we show only one particular set of parameters for
each case rather than trying to determine the complete allowed region in
parameter space.  For the same reason, we do not consider higher-order
corrections from MSSM thresholds \cite{Chun:1999vb,Chankowski:2001mx} or two-loop
effects \cite{Antusch:2002ek}, for example, which could change the
low-energy value of the solar angle and may make it necessary to adjust
some of the initial values, but do not spoil the general viability of
the scenario.

\subsection{Zeros in the 12- and 13-Entries}
The texture with $m_{\nu_{12}}=m_{\nu_{13}}=0$ (pattern $\mathrm{F}_1$ in the
classification of \cite{Guo:2002ei}) implies either $m_1=m_2=m_3$ or
$\theta_{12}=0$ and is therefore not compatible with data unless these
predictions are changed by quantum corrections.  Both cases have been
studied in the literature \cite{Chankowski:2000fp,Babu:2002dz,Antusch:2002fr}, and
it has been found that both can be compatible with the LMA solution at
low energy. Therefore, we do not give an example here.

\subsection{Vanishing Neutrino Mixings at High Energy}

\begin{figure}
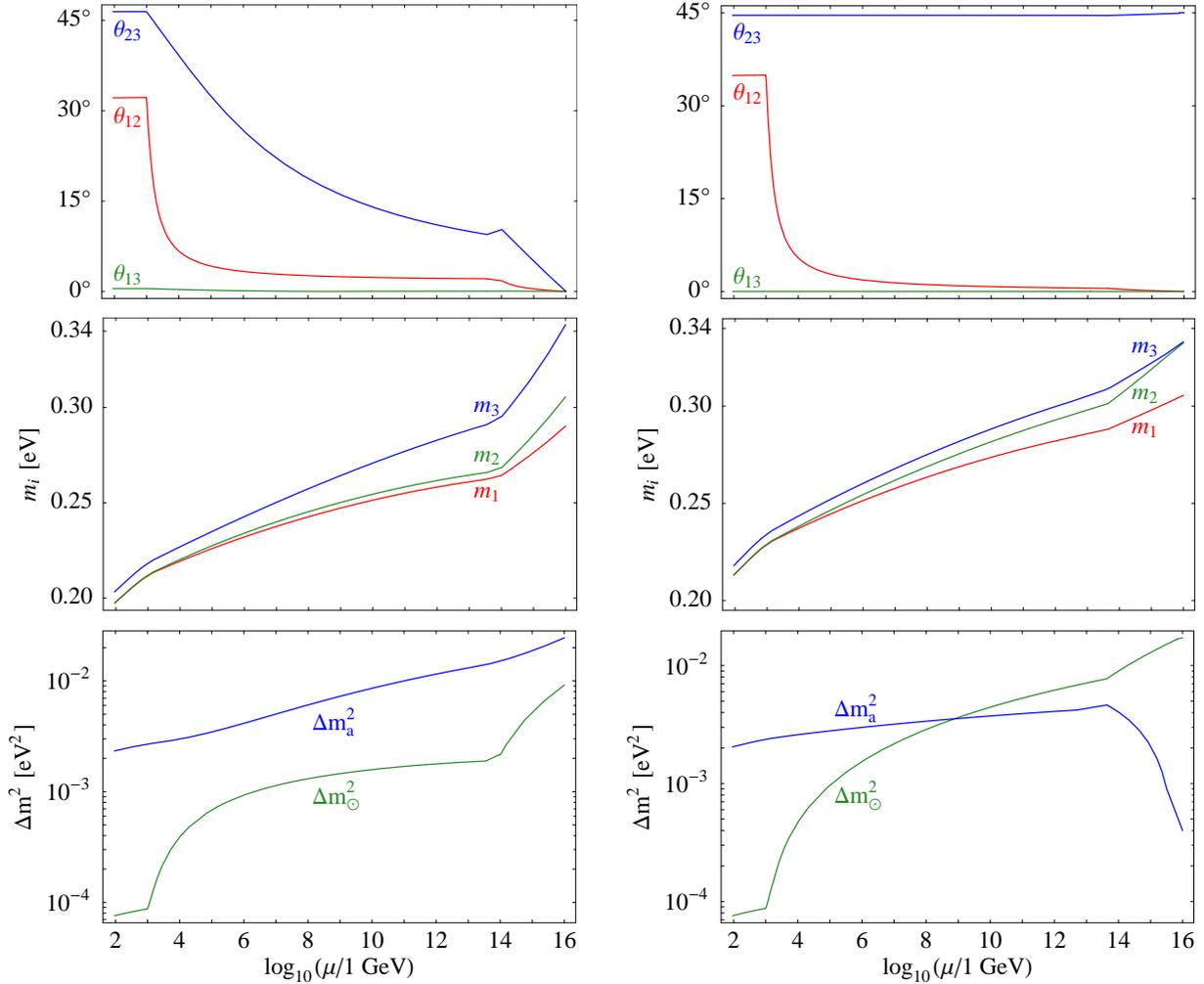

\centering
\begin{tabular*}{\textwidth}{@{}r@{\extracolsep\fill}r@{}}
\includegraphics[bb=80 618 318 743,clip,width=0.48\textwidth]{RevConstrDiagonalMAngles}
&
\includegraphics[bb=82 618 316 743,clip,width=0.48\textwidth]{RevConstrZeroM331213Angles}
\\
\includegraphics[bb=87 618 326 743,clip,width=0.48\textwidth]{RevConstrDiagonalMMasses}
&
\includegraphics[bb=88 618 325 743,clip,width=0.48\textwidth]{RevConstrZeroM331213Masses}
\\[2pt]
\includegraphics[bb=90 595 328 742,clip,width=0.48\textwidth]{RevConstrDiagonalMDms}
&
\includegraphics[bb=90 595 328 742,clip,width=0.48\textwidth]{RevConstrZeroM331213Dms}
\end{tabular*}
\caption{Running from vanishing neutrino mixing (left column) or a
three-zero texture with $m_{\nu_{33}}=m_{\nu_{12}}=m_{\nu_{13}}=0$
(right column) at the GUT scale $M_\mathrm{GUT}=10^{16}$ GeV to the LMA
solution at low energy.  We used the MSSM with $\tan\beta=50$ and a SUSY
breaking scale $M_\mathrm{SUSY}=1$ TeV.  The kinks in the plots stem
from integrating out the heavy singlets and the SUSY particles at their 
mass thresholds and the SUSY breaking scale, respectively.  The mass
eigenvalues are defined using the constant vacuum expectation value
$v = v(0) \approx \sin\beta \cdot 246$~GeV.
}
\label{fig:RevConstrZeros}
\end{figure}

The numerical analysis of cases where a vanishing 23-element is 
generated shows that the 12- and 13-elements often become very small, 
too. This prompts one to ask whether exactly vanishing mixings might 
be possible as well. As shown in the left column of 
\Figref{fig:RevConstrZeros}, this is indeed the case.  In this
example, we choose a diagonal neutrino mass matrix at
$M_\mathrm{GUT}=10^{16}$ GeV and use the MSSM with $\tan\beta=50$ and a
SUSY breaking scale of $M_\mathrm{SUSY}=1$ TeV, below which the SM
is assumed to be
valid as an effective theory.  The mass of the lightest neutrino at low
energy is about 0.2 eV.  The Majorana phases are set to zero without
loss of generality, since they can be absorbed by a redefinition of the
neutrino fields for a diagonal mass matrix.  The
numerical values of the light neutrino masses and Yukawa couplings at
the GUT scale are
\[
        m_\nu(M_\mathrm{GUT}) = {\small\begin{pmatrix}
         0.2902 & 0 & 0 \\ 0 & 0.3056 & 0 \\ 0 & 0 & 0.3434 
        \end{pmatrix}} \mathrm{eV},
\quad
        Y_\nu(M_\mathrm{GUT}) = {\small\begin{pmatrix}
         -0.04 & 0 & 0 \\ -0.01 & 0.71 & -0.02 \\ 0.004 & -0.37 & 0.93
        \end{pmatrix}} .
\]

The masses of the singlet neutrinos are 
$M_1 \approx 2 \cdot 10^{11}$ GeV, $M_2 \approx 3 \cdot 10^{13}$ GeV, 
and $M_3 \approx 10^{14}$ GeV.
At $M_Z$, we obtain $\theta_{12} \approx 32^\circ$, $\theta_{13} \approx
0.5^\circ$, $\theta_{23} \approx 46^\circ$, $\delta=0$, $\Dmsol \approx
7.6 \cdot 10^{-5}$ eV$^2$, and $\Dmatm \approx 2.3 \cdot 10^{-3}$
eV$^2$, all well within the experimentally allowed $2\sigma$ ranges.
For the neutrino mass matrix at the lowest see-saw scale, we find
\begin{equation}
        m_\nu(M_1) = {\small\begin{pmatrix}
        0.2558 & 1.2 \cdot 10^{-4} & -4.0 \cdot 10^{-5} \\
        1.2 \cdot 10^{-4} & 0.2600 & 0.0041 \\
        -4.0 \cdot 10^{-5} & 0.0041 & 0.2774
        \end{pmatrix}} \mathrm{eV} \;.
\end{equation}
It clearly satisfies the requirement of a strong hierarchy between
those elements that vanish at the GUT scale and the other entries, which
we derived in \Secref{sec:MassMatrixElements}.  Especially the large
$\theta_{23}$ at low energies is easier to obtain in the MSSM
than in the SM, since in the former the contribution of the
running below the see-saw scale is also significant, if $\tan\beta$ is
not too small, and drives the angle in the desired direction.
Nevertheless, working examples can be found in the SM as well.
Obviously, this result implies that the two-zero textures with
$m_{\nu_{12}}=m_{\nu_{23}}=0$ and $m_{\nu_{13}}=m_{\nu_{23}}=0$ are
possible as well.

One can see from the numerical precision of the above $m_\nu$ and
$Y_\nu$ at the GUT scale that some tuning is necessary
to obtain acceptable results.  However, this is not that surprising
considering that one needs an $\mathcal{O}(10^{-5}$ eV$^2)$ mass squared
difference between mass eigenvalues whose squares are of the order of
$10^{-2}$ eV$^2$.  Furthermore, there is a lot of freedom to rearrange
the values of the Yukawa couplings.

\subsection{Zeros in the 33-, 12- and 13-Entries}
Finally, we consider an example where both Majorana phases equal $\pi$.
As we derived in \Secref{sec:ConditionsForStability}, it should be
possible to create zeros in the entries $m_{\nu_{22}}$, $m_{\nu_{33}}$,
$m_{\nu_{12}}$, or $m_{\nu_{13}}$ in this case.  The
two-zero textures involving these elements (cases B and C in the
classification of \cite{Frampton:2002yf,Guo:2002ei}) are compatible with
data anyway.  Therefore, we pursue the more ambitious goal of creating a
three-zero texture with $m_{\nu_{33}}=m_{\nu_{12}}=m_{\nu_{13}}=0$.
Again, we use the MSSM with $\tan\beta=50$ and $M_\mathrm{SUSY}=1$ TeV.
The mass of the lightest neutrino at low energy is about 0.21 eV.  The
light neutrino mass matrix and the Yukawa couplings at the GUT scale are
\[
        m_\nu(M_\mathrm{GUT}) = {\small\begin{pmatrix}
         -0.30554 & 0 & 0 \\ 0 & -0.0006 & 0.3328 \\ 0 & 0.3328 & 0
        \end{pmatrix}} \mathrm{eV},
\quad
        Y_\nu(M_\mathrm{GUT}) = {\small\begin{pmatrix}
         -0.04 & 0 & -0.05 \\ 0.01 & 0.27 & 0 \\ 0.01 & -0.35 & 0.61
        \end{pmatrix}} .
\]
The singlet neutrinos have the masses 
$M_1 \approx 10^{11}$ GeV, $M_2 \approx 5 \cdot 10^{12}$ GeV, 
and $M_3 \approx 4 \cdot 10^{13}$ GeV.
For the oscillation parameters at low energy, we find 
$\theta_{12} \approx 35^\circ$, $\theta_{13} \approx 0.01^\circ$, 
$\theta_{23} \approx 45^\circ$, $\delta=0$, 
$\Dmsol \approx 7.6 \cdot 10^{-5}$ eV$^2$, and 
$\Dmatm \approx 2.0 \cdot 10^{-3}$ eV$^2$.
Their running is displayed in the right column of
\Figref{fig:RevConstrZeros}.  In this
example, the value of $\theta_{23}$ remains approximately $45^\circ$ at
all energies, since its RG evolution is strongly damped by the Majorana
phases.  At high energies, the mass ordering is inverted, but it changes
to a normal hierarchy because the running of $\Dmsol$ is very different
from that of $\Dmatm$.
Of course, it is also possible to create a zero $m_{\nu_{22}}$ instead
of $m_{\nu_{33}}$.  However, both diagonal elements can vanish at the
same time only if $m_{\nu_{12}}$ or $m_{\nu_{13}}$ remains finite.

\section{Discussion and Conclusions}
In this paper, the stability of zeros in neutrino mass matrices under 
quantum corrections has been studied and the consequences for the 
compatibility with experimental data have been discussed. 
We have considered a see-saw scenario with heavy singlet 
neutrinos where texture zeros in the effective mass matrix of the 
light neutrinos are assumed to be explained at the GUT scale. 
The discussion was performed in the basis where the mass matrix 
of the charged leptons is diagonal.  
The contributions of the off-diagonal elements of the neutrino 
Yukawa couplings to the RG running of the 
neutrino mass matrix can then replace the texture zeros by non-vanishing
entries.
From a bottom-up perspective, we have called this radiative
generation of texture zeros. The positions where
texture zeros can be generated depend on the CP parities of the mass
eigenstates. As a consequence of this analysis, we find that some 
textures for the neutrino mass matrix that have been classified 
as incompatible with experimental data are not excluded. We find 
that three of the six forbidden two-zero textures as well as 
several three-zero textures are allowed. We have also shown that 
the radiative generation of texture zeros is not possible for 
hierarchical neutrino masses or for Majorana phases significantly different 
from 0 and $\pi$. Hence, the usual classification of forbidden 
and allowed textures applies in these cases. Texture zeros in 
the 12- and 13-entries of the mass matrix can also be created for 
arbitrary Majorana phases as long as they are approximately equal. 
The validity of our results was demonstrated by numerical examples.
We included even a case where acceptable low-energy mixings are 
generated starting from a fully diagonal effective neutrino mass 
matrix at the GUT scale.

One should keep in mind that the mass matrix of the light neutrinos 
at high energy scales is a secondary quantity derived from the 
neutrino Yukawa couplings and the Majorana mass matrix of the singlet 
neutrinos. A more complete theory of flavour with texture zeros 
would probably be based on some symmetry predicting zeros in 
these matrices. As there are many possibilities, see e.g.\
\cite{Kageyama:2002zw,Bando:2003ei,Zhou:2004wz}, the analysis is
beyond the scope of this work and remains to be done in future
studies.  However, it is clear that a statement which was found to be
stable under radiative corrections in the present context will not
change as it does not depend on the origin of the zeros in the neutrino
mass matrix. A change is more likely the other way round: 
as the requirement of texture zeros in the Yukawa couplings and 
Majorana mass matrix removes some free parameters, the radiative 
generation of texture zeros becomes harder.  
It does not necessarily become impossible,
however, since zeros in these matrices can be generated
radiatively as well.  It is also possible that a flavour symmetry
guarantees the changes in the positions of the zeros below the GUT scale
to be small.  This obviously depends on the details of the model.

Another aspect concerns flavour-violating decays of charged leptons 
such as $\mu \to e\gamma$.  It should be possible to suppress these 
processes sufficiently by adjusting the Yukawa couplings in such a 
way that the relevant elements of $Y_\nu^\dagger Y_\nu$ are very small.

Finally, one could also ask if an originally allowed texture can become
forbidden at low energy.  In particular examples, this is certainly
possible, but it can always be avoided by choosing smaller Yukawa
couplings, a smaller mass of the lightest neutrino or different values
of the Majorana phases, all of which are not very strongly restricted
by experiment.  Hence, a complete exclusion of an allowed texture is
not possible.  In other words, the RG evolution can
make a forbidden texture allowed, but not vice versa.

\section*{Acknowledgments}
This work was supported in part by the 
``Sonderforschungsbereich~375 f\"ur Astro-{}Teil\-chen\-phy\-sik der 
Deutschen Forschungsgemeinschaft''.
We would like to thank Stefan Antusch, Jisuke Kubo and Michael Ratz for
interesting discussions and useful comments.

\bibliography{Running}
\bibliographystyle{TitleAndArxiv}

\end{document}